\documentstyle[12pt,epsf]{article}
\newcommand{\beq}{\begin{equation}}
\newcommand{\eeq}{\end{equation}}

\begin{document}

\title{Branes in external field or more about Randall-Sundrum scenario}
\author{K.Selivanov}
\date{}
\maketitle

\begin{abstract}
Fate of branes in external fields is reviewed with emphasis on a 
spontaneous creation of the Brane World. No negative tension brane
is involved. 
\end{abstract}

\section{Charged Branes}
The word "brane" in this talk will stand for a multidimensional 
relativistic massive film, possibly charged with respect to an 
$n$-form (gauge) field. The branes below can be viewed on as the fundamental
ones or as effective ones, e.g. domain walls.
Its action consists of two pieces, a tension term and a charge term.

Tension term reads
\begin{equation}
\label{tension}
S_{tension}=T \oint \sqrt {det({\hat g_{ind}})}
\end{equation}
the integration is over the brane world-volume, 
$g_{ind}$ is the metric induced on the
world-volume via its embedding into
the target space and the coefficient $T$ is the tension of the brane.
This term is an analog of the mass term for a particle,
$m \oint \sqrt{det({\hat g_{ind}})}$,
where integration is over world-line of the particle.\\
Charge term reads
\begin{equation}
\label{charge}
S_{charge}=e\oint {B},
\end{equation}
where ${B}$ is a $n$-form gauge field field,
corresponding curvature being $H=dB$.
Branes are sources for this field and they are affected
by this field.
In what follows we assume $H$ to be a top degree form.

Everything I am going to talk about is related to the 
Schwinger type process - production of branes by homogeneous
external field $H$.

\section{A warm up example}
Consider first a warm up example: production of particles by a 
homogeneous $E$ field.

For this case the effective action reads
\beq
\label{action}
S_{eff}=TL-eEA
\eeq
where $L$ is the length of the  world-line of the particles produced,
$T$ is a mass  of the particle,
(as usual, particle-antiparticle history looks like  a closed world-line),
and  $A$ is the area surrounded by the world-line. 

Notice, by the way, that
upon  appropriate identification
of the parameters, the same effective action describes
false vacuum decay in (1+1) scalar field theory.  Particles are 
substituted by kinks, electric field - by energy difference between 
false and true vacuum.

Extremal world-line for the action Eq.(\ref{action}) is a circle
of the radius 
${\bar r}=\frac{T}{eE}$.

\begin{figure}
\hspace*{1.3cm}
\epsfxsize=5cm
\epsfbox{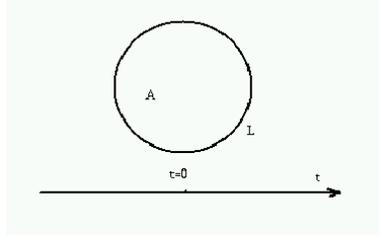}
\caption{Euclidean bounce for particle production.}
\end{figure}

Value of the effective action on this extremal curve (bounce) defines
the probability P of the spontaneous process:
follows
\beq
P \propto exp(-\frac{\pi T^2}{eE}).
\eeq

\begin{figure}
\hspace*{1.3cm}
\epsfxsize=5cm
\epsfbox{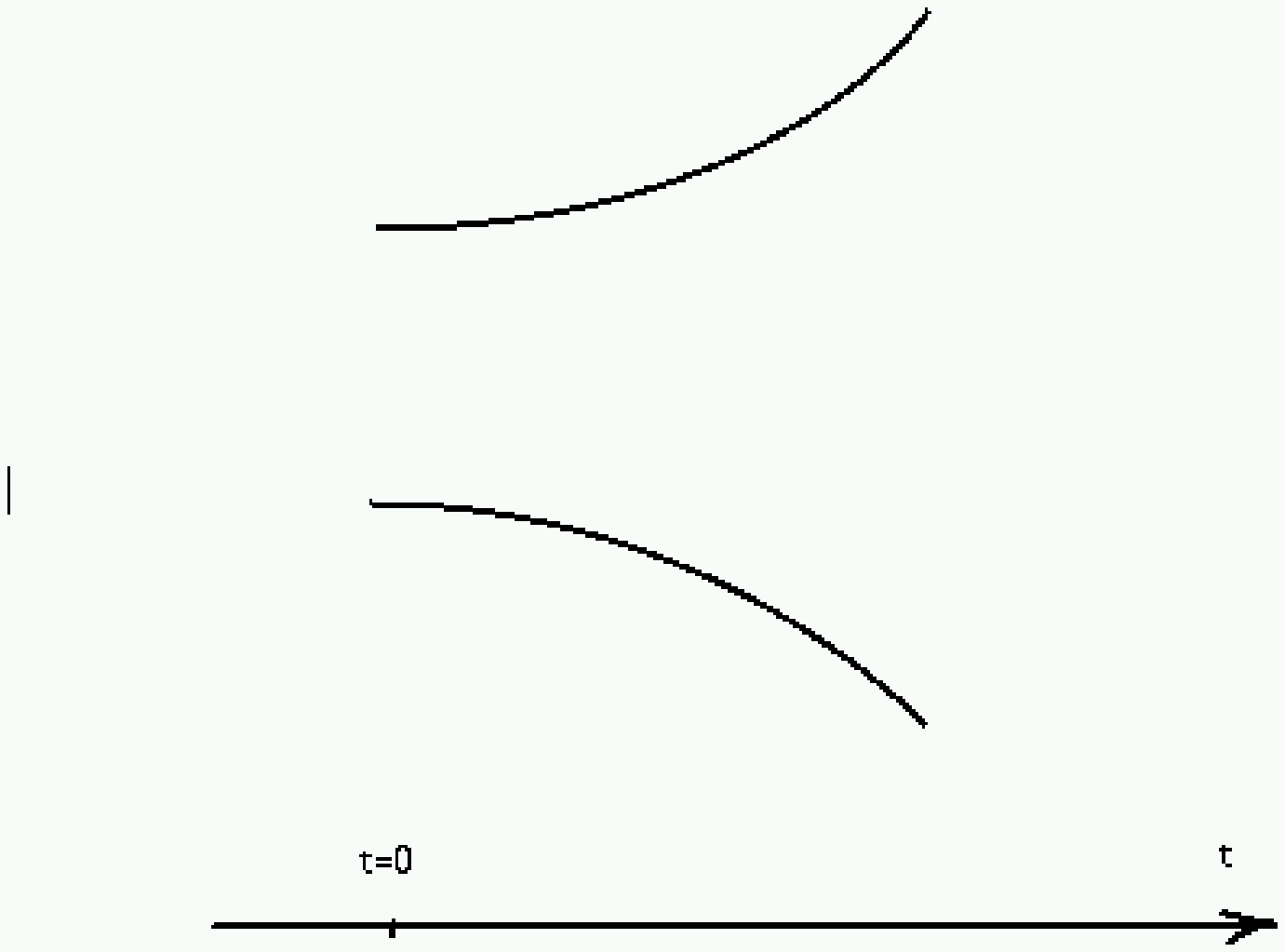}
\caption{Minkowski evolution of the particles produced.}
\end{figure}
Notice that Minkowski evolution is obtained from the Euclidean bounce 
by analitical continuation.

\section{Spontaneous production of branes}
The above warm-up example is easily generalized to the case
of spontaneous production of branes in a homogeneous $H$ field
\cite{bt}. The effective action
in the assumption of homogeneouty of $H$ reads
\beq
\label{actione}
S_{eff}=TA-ehV.
\eeq
World-volume of the brane  produced form a closed hypersurface
(like the world-line of the particle produced in the above example).
$A$ in Eq.(\ref{actione}) is the area of the world-volume,
$V$ - volume of the region inside the brane.
The  Euclidean bounce is a $p+1$-dimensional sphere of radius
$\bar{r}=\frac{(d-1)T}{eh}$.
The value of the effective action on the bounce defines probability
of the brane production:
\beq
P \propto exp(-const \frac{T^{d}}{(eh)^{d-1}}).
\eeq
 
Analogously to the above example, The Minkowski evolution can be obtained
from the Euclidean bounce via analytical continuation.

\begin{figure}
\hspace*{1.3cm}
\epsfxsize=5cm
\epsfbox{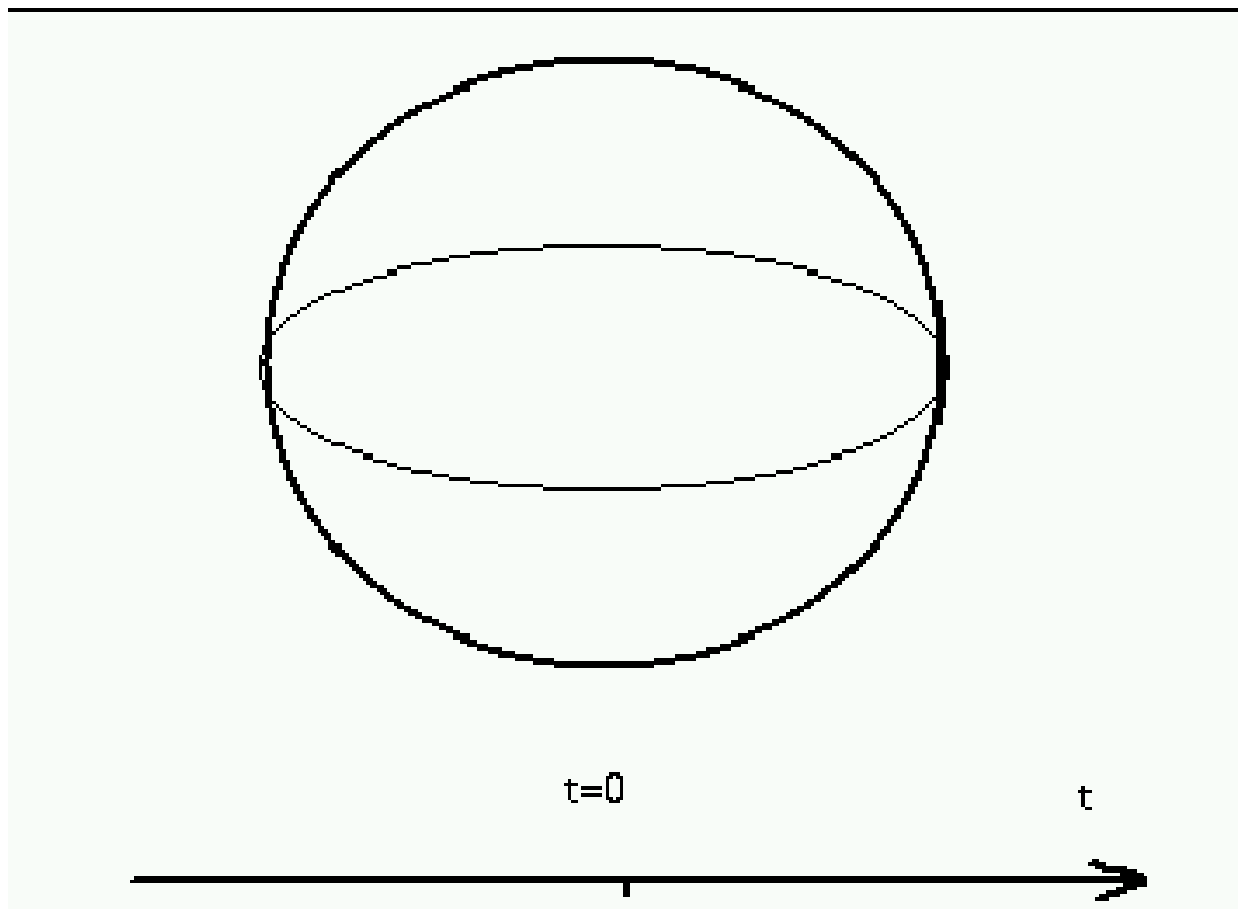}
\caption{Euclidean bounce for brane production.}
\end{figure}

\begin{figure}
\hspace*{1.3cm}
\epsfxsize=5cm
\epsfbox{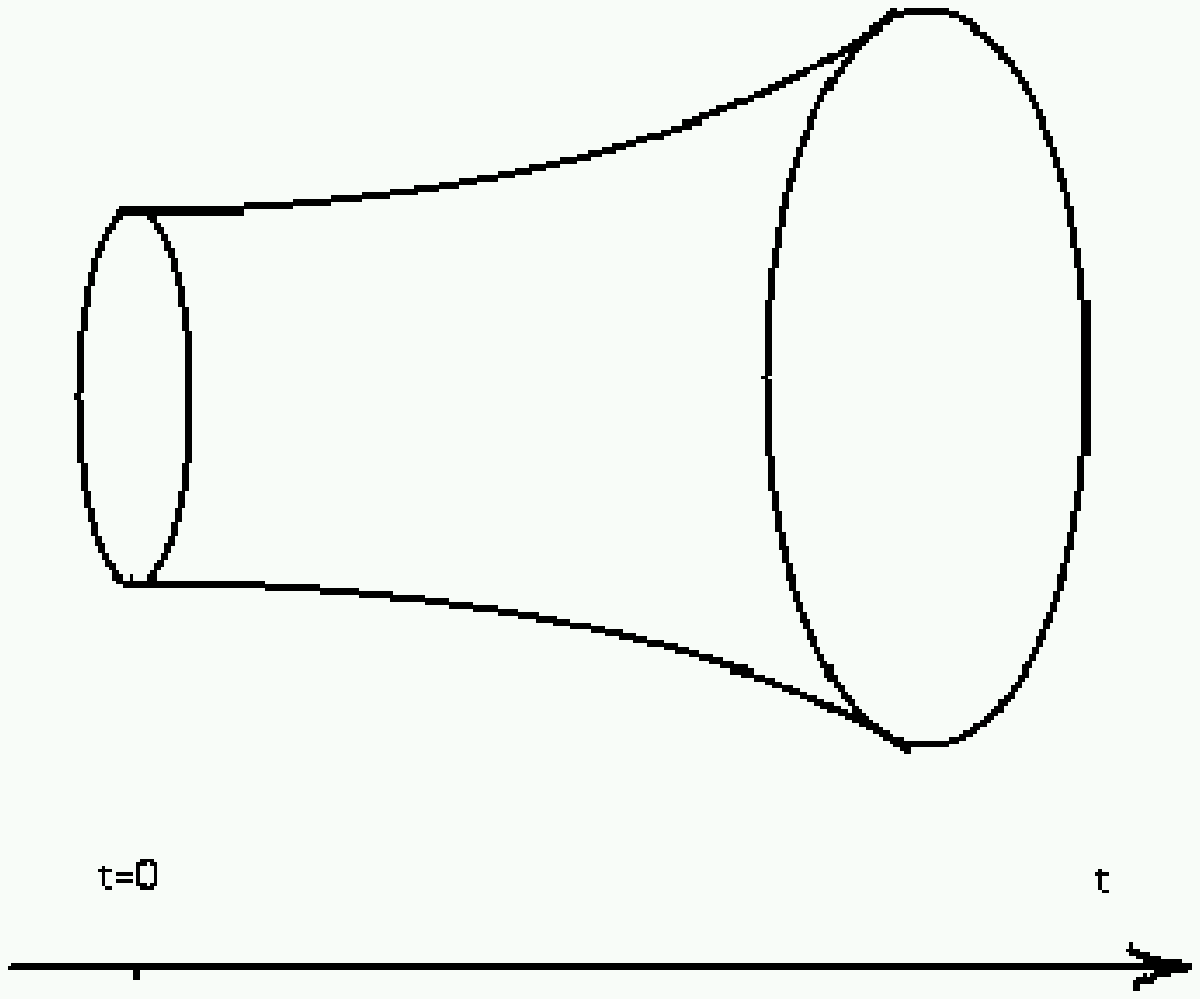}
\caption{ Minkowski evolution of the brane produced.}
\end{figure}

For branes it is important to include gravity, which also was done in
\cite{bt} and which I'll be back later on in my talk.

\section{Induced brane production} 
Let us now consider the induced brane production \cite{gs1}.
Before doing so, I need to introduce a new ingredient - brane junctions
\cite{john}. Brane world-volumes can meet. The manifold along which the 
world-volumes meet is called the junction manifold. Angles at which
the branes meet each other are fixed by the tension force balance 
condition. And, of course, charge of the branes is conserved.
That was about junctions.

The setup for the induced brane production is as follows. There is
external homogeneous field and there is external neutral brane which can
have junctions with charged branes to be produced. The question is
what is the probability.

Let us again begin with a warm up example - one particle induced decay in
(1+1)d \cite{sv}. The setup is as follows. On has a false vacuum and a 
massive particle 
in it such that the correspondig field has a zero mode on the wall which
could separate false and true vacuum. Then, if a bubble of the true vacuum
vacuum inside the false vacuum is produced, it is more profitable for the 
particle to ride a part of its way in the form of zero mode on the wall
of the bubble. Th effective action describing this case,
\beq
\label{actioni}
S_{eff}=mL-eEA + m {\tilde L},
\eeq
differs from Eq.(\ref{action}) only by the last term. $m$ is the mass of the 
extra particle, ${\tilde L}$ is the length of the extra particle world-line
(only outside the bubble).

\begin{figure}
\hspace*{1.3cm}
\epsfxsize=5cm
\epsfbox{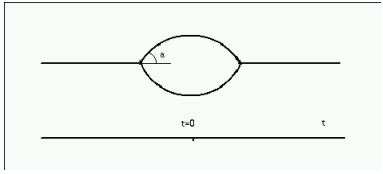}
\caption{Bounce for the induced vacuum decay.}
\end{figure}

The bounce is now glued of two segments of a circle of the same radius as
in the case of spontaneous decay. These two segments
meet  with the world-line of the particle (junction!) at the angle $\alpha$ 
which is defined by the force balance condition,
$m=2T \cos \alpha$.

The "charge conservation" in the present case is equivalent to a trivial fact 
that when one passes through the bubble crossing its wall twice, one gets 
back to false vacuum.
One then straightforwardly compute the probability of the false vacuum decay
\cite{sv}. There are two clear limiting cases. When mass of the particle is small
compared to the mass of the wall, bounce is not disturted and probability is the 
same in spontaneous case. When mass of the particle is close to $2T$, bounce shrinks 
to a point and there is no exponential suppression in the induced vacuum decay.

Generalization to the case of branes is more or less obvious \cite{gs1}.

\begin{figure}
\hspace*{1.3cm}
\epsfxsize=5cm
\epsfbox{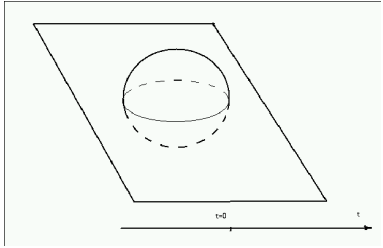}
\caption{ Bounce for the induced brane production.}
\end{figure}

The bounce now consists of two segments of (d-1) dimensional sphere of the
same radius as in the case of spontaneous brane production,
glued to the external brane along the junction manifold. The force balance 
condition reads
${\tilde T}=2T \cos \alpha$, 
where ${\tilde T}$ is the tension of the external neutral brane.
Minkowski evolution is again obtained by analytical continuation of the bounce.

\begin{figure}
\hspace*{1.3cm}
\epsfxsize=5cm
\epsfbox{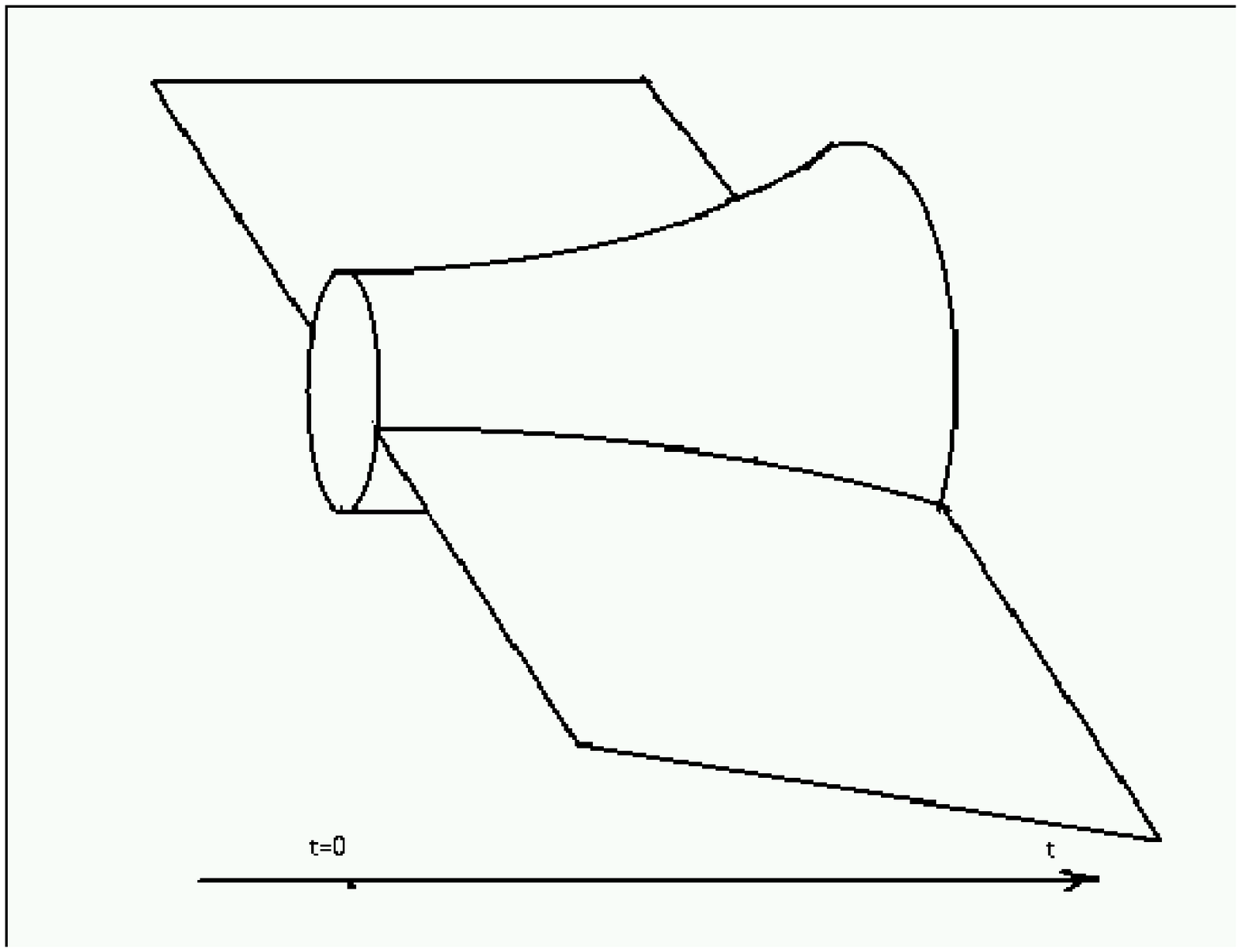}
\caption{ Minkowski evolution for the induced brane production.}
\end{figure}

Calculation of the probability is straightforward, two clear limiting
cases are as in the warm up example.

\section{A sketch of Brane World}
Now I would like to make a digression to sketch an idea of the Brane World.
This is an alternative to the idea of compactification. In compactification
extra dimensions are compact and small and thus cannot be seen at moderate energies.
In the brane world, extra dimensions are infinite, but the matter \cite{rubshap},
gauge fields \cite{pw} and gravity \cite{rs} are localized on a brane
(domain wall or other topological defect in the extra dimensions).

The model considered in \cite{rs} included two branes localized at different
points on a circle (the 5th dimension was taken $S^{1}$ of arbitrarily large radius).
One of the branes was physical (RS-brane), the other was a so-called  
regulator brane (R-brane). The gravity was localized on the RS brane. The drawback
of the model was that R-brane had a negative tension (see e.g. discussion in
\cite{witten}).

Many modifications of the construction in \cite{rs} were studied
(see e.g. \cite{more} and other references to the original paper
\cite{rs}), most of them included the negative tension brane.

Let us consider in more detail the construction in \cite{grs}. 

\begin{figure}
\hspace*{1.3cm}
\epsfxsize=5cm
\epsfbox{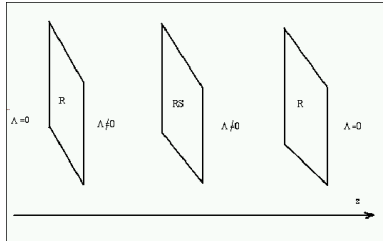}
\caption{ Gregory-Rubakov-Sibiryakov model.}
\end{figure}

It included three branes localized on a line (5th dimension was taken $R^{1}$),
two negative tension R-branes and one RS-brane between them. Cosmological constant
outside R-branes was zero, cosmological constant between R-branes and RS-brane was 
negative. Notice that in this model cosmological constant is not really a constant,
it jumps on the R-branes. So, in fact, there is a hidden external field in the model,
and R-branes are charged with respect to it. This motivates the following 
construction in \cite{gs2}.

\section{Spontaneous Brane World creation}
I would now like to describe spontaneous creation of the Brane world
in a homogeneous external field. The process considered is a sort of inverse
to the induced brane production in the external field. The bounce now consists
of two segments of the charged branes (R-branes) which are glued along the junction 
manifold with the neutral brane (RS-brane) which is inside the bubble.

\begin{figure}
\hspace*{1.3cm}
\epsfxsize=5cm
\epsfbox{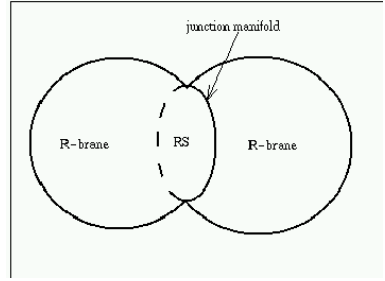}
\caption{ Tunneling into the Brane World.}
\end{figure}

The relevant effective action reads
\begin{eqnarray}
\label{action3}
S= \frac{1}{2k}\int d^5 x \sqrt g (-R + 2\Lambda)
+\frac{1}{k} \oint \sqrt g_{ind}K+\nonumber\\
\frac{1}{2} \int d^5 x \sqrt g h^2 +
T_{R} \oint_{R} \sqrt g_{ind} +
T_{RS} \oint_{RS} \sqrt g_{ind}
\end{eqnarray}
Let us explain ingredients in Eq.(\ref{action3}).
$h$ is Hodge dual scalar of the field strength $H$.
Being a top form, this field does not propagate.  Field equations
say that in empty space $h$ is a constant, in our case it only jumps
at the charged R-branes by its charge e,
$h_{+}- h_{-}=e$,
where $h_{+}$ is outside value of $h$, $h_{-}$ is its inside value.
In the effective action Eq.(\ref{action3}) it is assumed that
field equations for $h$ are resolved, the charge of the R-branes
enters effective action only via $h^2$-term.

Constant k in Eq.(\ref{action3}) is the five dimensional gravitational
constant, $R$ - scalar curvature, $\Lambda$ stands for cosmological constant,
for which we assume that it is negative and exactly compensates
energy density of the field $h$ outside,
$\Lambda + \frac{k h_{+}^2}{2} =0$
so that outside the metric is flat.  Notice right away that effective
cosmological constant  inside is
\beq
\label{cc}
\Lambda_{eff}=\Lambda + \frac{k h_{-}^2}{2}=
- \frac{k h_{+}^2}{2}+ \frac{k h_{-}^2}{2}.
\eeq
and hence the scalar curvature $R$ of the AdS metric inside
any of the two segments in figure reads
$R=\frac{2d}{d-2} \Lambda_{eff}=\frac{10}{3} \Lambda_{eff}$,
and the corresponding AdS radius, $R_{AdS}$, reads
$R^{2}_{AdS}=-\frac{2(d-1)(d-2)}{\Lambda_{eff}}=-\frac{24}{\Lambda_{eff}}$.

The origin of two of the three surface terms in Eq.(\ref{action3}) is obvious -
these are tension terms for R- and RS-branes. The third term,
$ \frac{1}{k} \oint \sqrt g_{ind}K$ is introduced to ensure
that variation of the curvature term does not depend on normal derivatives
of variation of the metric on the branes \cite{israel}.
Here $K=g_{ind}^{ij}K_{ij}$ stands for the trace of the external curvature
$K_{ij}$ of the branes, the integral is over all branes, every brane
contributing twice, with $K$ computed in metric on one or the other side of
the brane.

So much about ingredients in the effective action Eq.(\ref{action3}).
Let us now explain details of the bounce solution.
The metric to the right of the RS-brane inside R-brane   reads
\beq
\label{plus}
ds^{2}=\frac{dz^2+d\rho^2+\rho^2d\Omega^2_{3}}
{(1-\frac{(z-a)^2+\rho^2}{R^{2}_{AdS}})^2}
\eeq
while the metric to the left of the RS-brane inside R-brane   reads
\beq
\label{minus}
ds^{2}=\frac{dz^2+d\rho^2+\rho^2d\Omega^2_{3}}
{(1-\frac{(z+a)^2+\rho^2}{R^{2}_{AdS}})^2}
\eeq
where $z$ is coordinate along the axes of symmetry orthogonal to
RS-brane, $\rho$ is the radial coordinate in orthogonal to $z$
directions,  $d\Omega^2_{3}$ is the metric of the corresponding
3-sphere,  $a$ is a parameter.

The metrics Eq.(\ref{plus}),(\ref{minus}) are to be sewed
on R-branes with flat metric  and on RS-brane with
each other, in the sence that metrics themselves are continues,
while their normal derivatives jump according to the Israel
condition,
$\Delta K_{ij}=\frac{kT}{d-1}g_{ij}=\frac{kT}{4}g_{ij}$.
On R-branes this condition fixes radius ${\bar R}$
of the spherical segments,
${\bar R}=\frac{2T_{R}(d-1)}{h^{2}_{+}-h^{2}_{-}}$,
which is of course the same as for bounce without
RS-brane,
and on RS-brane this condition fixes the parameter
$a$,
$a=\frac{T_{RS}(d-1)}{h^{2}_{+}-h^{2}_{-}}$.
Importantly,
$cos{\alpha}=\frac{a}{\bar R}=\frac{T_{RS}}{2T_{R}}$,
which is precisely the force balance condition at the junctions.

Substituting these data into the effective action Eq.(\ref{action3})
one straigtforwardly obtains exponential factor for the probability
P of the process. It ranges between
$P\propto e^{-S_{bt}}$
for very light RS-brane,  $T_{RS}\ll T_{R}$,  where
$S_{bt}$ is the action for the bounce without RS-brane, which has been
computed in  \cite{bt},
and
$P\propto e^{-2S_{bt}}$
for  $T_{RS}=2 T_{R}$. More heavy RS-brane cannot be produced in this way.

Since the internal brane is located at $z=0$  the
induced metric of 4D world is immediately seen from Eq.(\ref{plus}).
It appears to be the
AdS space  with the AdS radius
$R_{AdS4}^2=R_{AdS}^2-a^2$
and, correspondingly, the cosmological constant
$\Lambda_4=-\frac{12}{(R_{AdS}^2 -a^2)}$

After continuation into Minkowski space our Universe is a spacially finite, 
growing slice of AdS space. Spacial radius of R-brane is also growing, and we
are eventually left with the picture of \cite{rs}, or \cite{grs}, type.

\section{Some other related solution} 
I would also like to describe some other related solution obtained 
in\cite{gs2}.

Interestingly, one can have multiple RS branes in the final state.
In that case we have additional pair of the R brane segments
of the same radius $\bar R$ for each new RS  brane.
The metric between
the n-th and n+1 -th RS branes reads
\beq
ds^{2}_{n}=\frac{dz^2+d\rho^2+\rho^2d\Omega^2_{3}}
{(1-\frac{(z-a_{n})^2+\rho^2}{R^{2}_{AdS}})^2}
\eeq
where $a_{n}=(2n-1)a$.
We choose z=0 at the center of the left R brane segment.

The distance
between branes in fifth dimension is $\delta x_{5}=2a$.
Let us interpret the RS branes as D branes which are neutral
with respect to the external NS field.
The picture described would amount to the
generic U(N) gauge group on their worldvolume.
Let us note  that since the distance between branes can be
interpreted as the vacuum expectation values of the scalar
field on the worldvolume of D branes  we have no
moduli associated to scalars in this solution.

\section{Conclusion}
My conclusions are as follows:\\
1. We have described induced brane production in external field;\\
2. We have described tunneling into the Brane World (a sort of Big Bang);\\
3. No negative tension branes;\\
4. 5d early Universe;\\
5. non-flat (AdS) 4d brane - curable (see \cite{gs3}).\\

I am thankful to A.Gorsky for the fruitful collaboration.


\begin{thebibliography}{99}  

\bibitem{bt}
J.~D.~Brown and C.~Teitelboim,
Nucl.\ Phys.\  {\bf B297}, 787 (1988).
\bibitem{gs1}
A.~Gorsky and K.~Selivanov,
Nucl.\ Phys.\  {\bf B571}, 120 (2000)
[hep-th/9904041].
\bibitem{john}
J.~Schwarz, Nucl. Phys. (Proc. Suppl) {\bf B55} (1997).
\bibitem{sv}
K.~G.~Selivanov and M.~B.~Voloshin,
JETP Lett.\  {\bf 42}, 422 (1985).
\bibitem{rubshap}
V.~A.~Rubakov and M.~E.~Shaposhnikov,
Phys.\ Lett.\  {\bf B125} (1983) 136.\\
K.~Akama,``An early proposal of 'brane world',''Lecture Notes in Physics, 176, (Springer-Verlag, 1983), 267-271, hep-th/0001113.
\bibitem{pw}\ J.~Polchinski, Phys. Rev. Lett. {\bf 75} (1995),4724 \\
E.~Witten, Nucl. Phys. {\bf B460} (1996) 335
\bibitem{rs}
L.~Randall and R.~Sundrum,
Phys.\ Rev.\ Lett.\  {\bf 83}, 4690 (1999)
[hep-th/9906064].
\bibitem{witten}\ E.~Witten,hep-ph/000229.
\bibitem{more}I.~I.~Kogan, S.~Mouslopoulos, A.~Papazoglou, G.~G.~Ross and J.~Santiago,
hep-ph/9912552.\\I.~I.~Kogan and G.~G.~Ross,hep-th/0003074. \\G.~Dvali, G.~Gabadadze and M.~Porrati,hep-th/0002190 \\C.~Csaki, J.~Erlich and T.~J.~Hollowood,hep-th/0002161.
\bibitem{grs}
R.~Gregory, V.~A.~Rubakov and S.~M.~Sibiryakov,
hep-th/0002072.
\bibitem{gs2}A.~Gorsky and K.~Selivanov, Phys.Lett.B485 (2000) 271, 
hep-th/0005066.
\bibitem{israel}
W.~Israel,
Nuovo Cim. 44B (1966) 1; 48B (1967) 463
\bibitem{gs3}A.~Gorsky and K.~Selivanov, hep-th/0006044


\end{thebibliography}
\end{document}